\documentclass[3p, numer, sort&compress, 12pt, review]{elsarticle}

\usepackage{graphicx,amsmath,amssymb,xcolor,wasysym}
\usepackage{subcaption}
\newcommand{\uu}[1]{\ensuremath{\,\mathrm{#1}}}

\newcommand{\figwidth}{8cm}
\newcommand{\smallfigwidth}{5.3cm}
\let\onlinecite\cite

\begin{document}

\title{Size-dependent surface energies of Au nanoparticles}

\author[DMW]{D. Holec\corref{corauthor}}
\ead{david.holec@unileoben.ac.at}
\author[DMW]{P. Dumitraschkewitz\fnref{current}}
\author[IM]{F.D. Fischer}
\author[Nano]{D. Vollath}
\address[DMW]{Department of Physical Metallurgy and Materials Testing, Montanuniversit\"at Leoben, Franz Josef Stra\ss{}e 18, A-8700 Leoben, Austria}
\address[IM]{Institute of Mechanics, Montanuniversit\"at Leoben, Franz Josef Stra\ss{}e 18, A-8700 Leoben, Austria}
\address[Nano]{NanoConsulting, Primelweg 3, D-76297 Stutensee, Germany}
\cortext[corauthor]{Corresponding author}
\fntext[current]{Now at Chair of Nonferrous Metallurgy, Department of Metallurgy, Montanuniversit\"at Leoben,Franz-Josef-Str. 18, Leoben 8700, Austria}
\date{\today}

\begin{abstract}
Motivated by often contradictory literature reports on size dependence of surface energy of gold nanoparticles, we performed  an atomistic study combining molecular dynamics and \textit{ab initio} calculations. 
We show that in the case of Au nanocubes, their surface energy converges to a value for $(0\,0\,1)$ facets of bulk crystals. 
A fast convergence to a single valued surface energy is predicted also for nanosheres. 
In this case, however, the value of the surface energy is larger than that of any low-index surface facet of bulk Au crystal. 
This fact can be explained by the complex structure of the surface with an extensive number of broken bonds due to edge and corner atoms. 
A similar trend was obtained also for the case of cuboctahedron-shaped nanoobjects. 
As the exact surface area of the nanoparticles is an ill-defined quantity, we introduced the surface-induced excess energy and discussed this quantity as a function of (i) number of atoms forming the nanoobject or (ii) the nanoobject characteristic size. 
In case (i), a universal power-law behaviour was obtained independent of the nanoparticle shape.
Importantly, we show that the size-dependence of the surface is hugely reduced is the surface area correction due to the extend of electronic cloud is considered, a phenomenon specifically important for small nanoparticles.
\end{abstract}

\begin{keyword}
  surface energy \sep nanoparticles \sep gold \sep ab initio \sep molecular dynamics

  \PACS
  31.15.A- \sep 
  61.46.Df \sep 
  68.35.Md \sep 
  83.10.Rs 
\end{keyword}

\journal{Applied Surface Science}

\maketitle

\section{Introduction}
Whenever one discusses thermodynamics or any other feature of nanoparticles, surface energy is one of the key properties.
This is due to the fact that volume-to-surface ratio is fast decreasing with decreasing nanoparticle size, hence enhancing influence of the surface properties. 
For example, it has been repeatedly reported that a modification of the surface energy leads to a corresponding change of the nanoparticle shape \cite{Chen2005-vn,Grzelczak2008-ji}. 
It is, therefore, surprising that for many materials this is a rather poorly known material property, especially when referring to small particles with characteristic dimension below $5\uu{nm}$. 
In many cases one finds decreasing surface energy with decreasing particle size, e.g., in a study by \citet{Vollath2014-vr} or earlier studies \cite{Safaei2010-rk,Attarian_Shandiz2008-sx}. 
On the one hand, this trend is conventionally explained with an increasing tendency to form a liquid-like structure at the surface of the particles \cite{Chang2005-fo}. 
On the other hand, there exists a series of primarily theoretical papers finding a significant increase of the surface energy with decreasing particle size, see, e.g., Refs. \onlinecite{Medasani2007-fb} or \onlinecite{Medasani2009-ir}. 
Furthermore, there are also some heavily disputed experimental results indicating an increasing surface energy with decreasing particle size \cite{Nanda2008-fd,Nanda2003-mh}. 
\citet{Nanda2003-mh} pointed out that the difference between various reported trends stems from the nanoparticle nature: for free nanoparticles, the surface energy is expected to increase with decreasing particle size, while the opposite trends is obtained for nanoparticles embedded in a matrix.
We note, however, that the derivation of the Kelvin equation as used in that work at elevated temperature is valid only at $T=0\uu{K}$.

Despite similar studies have been performed for other systems, e.g. Ag \cite{Medasani2007-fb, Medasani2009-ir}, there is only one report concerning specifically gold nanoparticles with various nanoparticle sizes \cite{Ali2015-mh}.
In our previous works \cite{Vollath2017-cn, Holec2017-qy}, only a cluster composed of 55 Au atoms was considered, leading to the conclusion that, as a consequence of the small nanoparicle size, the amorphous structure is the most preferable one even at $0\uu{K}$.
Nevertheless, the previous reports did not discuss the effect of nanoparticle shape in conjunction of their size. 
To unveil such trends we consider a model system of free nanoparticles with idealised shapes carved out of infinitely large bulk crystalline material, and subsequently structurally relaxed. 
This study addresses surface energy of gold crystalline nanoparticles, i.e., interfaces between the gold nanoparticles with well defined shape and vacuum. 
We note that this makes our results somewhat different from experiments, in which a solid nanoparicle--liquid solution interface or liquid-like layer at the nanoparicle surface may be of crucial importance during the forming process.

\section{Methodology}
Molecular dynamics (MD) simulations were performed using the LAMMPS package \cite{Plimpton1995-wi} together with an interatomic potential describing the gold--gold interaction within the embedded atom method (EAM) as parametrised by \citet{Grochola2005-tf}. 
The individual idealised nanoparticles with well-defined shapes were cut out from bulk fcc structure with lattice constants of $4.0694\,\mbox{\AA}$. 
This was obtained from fitting calculated total energies corresponding to different bulk volumes with Birch-Murnaghan equation of state \cite{Birch1947-nr}, and agrees well with the values $4.0701\,\mbox{\AA}$ obtained by \citet{Grochola2005-tf}. 
All structural models were relaxed using conjugate-gradient energy minimisation scheme with force-stopping convergence criterion set to $10^{-12}\,\mbox{eV/\AA}$. 
The rectangular simulation box was $\approx10\,\mbox{\AA}$ larger than the nanoparticles in order not to limit the relaxation procedure.
 
Additionally, a few \textit{ab initio} runs were performed to benchmark our MD calculations. We used Vienna Ab initio Simulation Package (VASP) \cite{Kresse1996-gt,Kresse1996-tg} implementation of Density Functional Theory (DFT)\cite{Hohenberg1964-in,Kohn1965-rd}. 
The plane wave cut-off energy was set to $400\,\mbox{eV}$, and the reciprocal space sampling was equivalent to $10\times10\times10$ $k$-mesh for the fcc-conventional cell. 
Two common approximations of the electronic exchange and correlation effects were considered: local density approximation (LDA) \cite{Kohn1965-rd} and the Perdew-Wang parametrisation of the generalised gradient approximation (GGA) \cite{Wang1991-ca}. 
The contribution of ions and core electrons were described by projector augmented wave (PAW) pseudopotentials \cite{Kresse1999-if}. 
Due to the employed periodic boundary conditions, we used a simulation box $\approx20\,\mbox{\AA}$ larger than the actual (unrelaxed) nanoparticle to avoid any undesired interactions through the vacuum separating neighbouring nanoparticles. 
Similarly, $\approx15\,\mbox{\AA}$ vacuum in the direction perpendicular to a free surface was used to separate slabs for calculating the surface energies of bulk Au (both for MD and DFT calculations).

\section{Results}

\subsection{Low-index facets of bulk Au}\label{sec:surface_energy_bulk}
The results presented in this chapter serve the subsequent discussion of the MD results, and their accuracy with respect to first principles calculations. 
Surface energy, $\gamma$, of surface facet $(h\,k\,l)$ can be calculated as
\begin{equation}
  \gamma=\frac1{2A}\left(E_{\text{slab}}-N E_{\text{bulk}}\right)\ ,
\end{equation}
where $E_{\text{slab}}$ is energy of a slab composed of $N$ layers. 
$E_{\text{bulk}}$ is the energy of the bulk material per one layer of cross-section $A$. 
The factor $2$ results from the fact that the slab has two surfaces. 
A layer is understood as a surface primitive cell, i.e. when the desired facet $(h\,k\,l)$ is perpendicular to one of the lattice vectors (for a detailed description of the surface primitive cells, see e.g., Ref.~\onlinecite{Holec2012-qs}). 
Due to the interaction of the two free surfaces, either through the vacuum (i.e., not well separated slabs in the case of periodic boundary conditions) or the bulk of the slab (i.e., too thin slab), the value $\gamma$ has to be converged with respect to both of these. 
In the case of MD simulations, only the latter convergence needs to be tested if the simulation is run in a box without periodic boundary conditions in the direction perpendicular to the free surface.

%

Test calculations revealed that vacuum of $10\uu{\AA}$ is sufficient to get surface energy results converged to well below $1\uu{meV/\AA^2}$. 
Similarly, a slab thickness of about $40\uu{\AA}$ is needed in order to avoid interactions of the free surfaces through the gold layer.
The obtained values from the DFT benchmarks and MD simulations are summarised in Table~\ref{tab:surface_energy}. 
The here obtained DFT values are comparable with data from the literature. 
They exhibit the same ordering ($\gamma_{(1\,1\,0)}>\gamma_{(1\,0\,0)}>\gamma_{(1\,1\,1)}$) as reported earlier \cite{Vitos1998-ix}. 
In a simplified picture, the surface energy expresses energy penalty related to the areal density of broken bonds \cite{Holec2012-qs,Kozeschnik2007-zs}. 
This is $8/a_0^2$ for the $(1\,0\,0)$ surface, $7.07/a_0^2$ for $(1\,1\,0)$, and $4.33/a_0^2$ for the $(1\,1\,1)$ surface ($a_0$ being the fcc lattice constant). 
The density of broken bonds is similar for the $(1\,0\,0)$ and $(1\,1\,0)$ surfaces, while it is significantly lower for the $(1\,1\,1)$ orientated facet, hence providing a qualitative explanation for the surface energy ordering.

\begin{table*}
\centering
\caption{Calculated surface energies for three low-index facets, including data from literature for comparison. (FCD = full charge density)}
\begin{tabular}{l|cccccc}
  & \multicolumn{2}{c}{$(1\,0\,0)$} & \multicolumn{2}{c}{$(1\,1\,0)$} & \multicolumn{2}{c}{$(1\,1\,1)$} \\
  & [$\text{meV}/\text{\AA}^2]$ & $[\text{J}/\text{m}^2]$ & [$\text{meV}/\text{\AA}^2]$ & $[\text{J}/\text{m}^2]$& [$\text{meV}/\text{\AA}^2]$ & $[\text{J}/\text{m}^2]$ \\\hline
  DFT-GGA (this work)& 54.5 & 0.87 & 57.0 & 0.91 & 45.2 & 0.72 \\
  DFT-GGA (Ref.~\onlinecite{Crljen2003-sn}) & & & & & 50 & 0.80 \\
  FCD-GGA (Ref.~\onlinecite{Vitos1998-ix}) & 101.5 & 1.63 & 106.1 & 1.70 & 80 & 1.28 \\

  MD (this work)& 80.9 & 1.30 &  &  & 72.5 & 1.16 \\

  DFT-LDA (this work)& 83.5 & 1.34 & 89.2& 1.43 & 78.4 & 1.26 \\
  DFT-LDA (Ref.~\onlinecite{Crljen2003-sn}) & & & & & 80 & 1.28 \\

  experiment (Ref.~\onlinecite{Galanakis2002-td}) & & & & & 93.6 & 1.50 \\
  experiment (Ref.~\onlinecite{Tyson1977-is}) & & & & & 94.0 & 1.51
\end{tabular}
\label{tab:surface_energy}
\end{table*}

The DFT and MD values exhibit an almost constant difference between the corresponding surface energies. 
Moreover, the MD values are very close to the DFT-LDA results. 
This is a somewhat surprising result since the EAM potential has been fitted to the DFT-GGA data using the same parametrisation by Perdew and Wang \cite{Wang1991-ca} as used here. 
We speculate that this is caused by fixing $4.07\,\mbox{\AA}$ as the lattice constant during the EAM potential fitting \cite{Grochola2005-tf}, as our LDA and GGA calculations yielded $4.061$ and $4.176\,\mbox{\AA}$, respectively. 
Nevertheless, since LDA and GGA are known to overestimate and underestimate, respectively, binding\cite{Haas2009-tu}, and since the MD values are in between the two DFT-based estimations, we conclude that the interatomic potential used here is suitable for studying trends in surface energies. 
Moreover, the resulting values are expected to be very close to DFT-LDA calculations.

\subsection{Impact of shape and size on the nanoparticles surface energy}\label{sec:surface_energy}

The surface energy of a gold nanoparticle consisting of $N$ atoms is defined as an excess energy with respect to the energy of $N$ atoms of bulk fcc gold, normalised to the nanoparticle surface area, $A$:
\begin{equation}
  \gamma=\frac{E_{\text{nanoparticle}}-N E_{\text{bulk}}}{A}\ .
  \label{eq:gamma}
\end{equation}
In the above, $E_{\text{nanoparticle}}$ is the total energy of the nanoparticle, while $E_{\text{bulk}}$ is energy per atom of bulk fcc Au. 
Unlike the total energies, the surface area $A$ is not a well defined quantity.
In the following sections, an area of a convex hull of the relaxed ionic positions is consistently used as an estimate for $A$.

\subsubsection{Nanocubes}
\begin{figure}
  \centering
    \includegraphics[height=\figwidth]{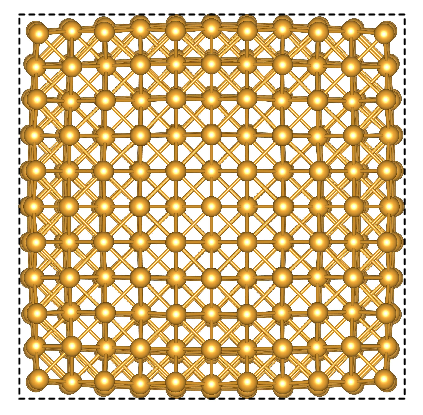}
  \caption{Relaxed structure of a nanocube with side $a=2.035\uu{nm}$ (666 atoms). 
           The dashed line is a guide for the eye showing an ideal square shape.}
  \label{fig:snapshot}
\end{figure}

In order to calculate the total energy of $\{1\,0\,0\}$-faceted nanocubes, structural models with a side length up to $20\uu{nm}$ were fully structurally relaxed. 
As a consequence of the surface tension, the apexes ``popped in'' as it is apparent from the snapshot of relaxed atomic positions shown in Fig.~\ref{fig:snapshot}.

\begin{figure}
  \centering2
  \includegraphics[height=\figwidth]{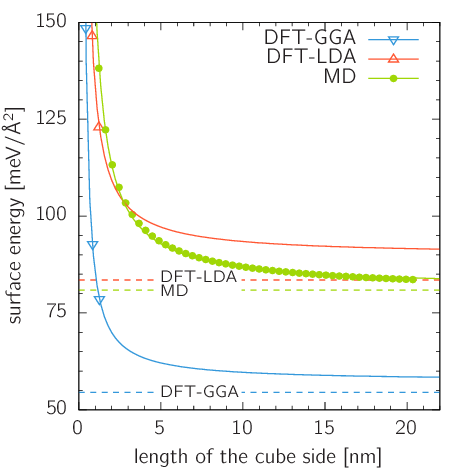}
  \caption{Surface energy of nanocubes calculated by DFT and MD.  
    The calculated datapoints were fitted with Eq.~\ref{eq:fit}. 
    The dashed lines are $(1\,0\,0)$ surface energies as listed in Table~\ref{tab:surface_energy}.}
  \label{fig:nanocubes}
\end{figure}

It was feasible to handle nanocubes only up to the $3\times3\times3$ supercell of the conventional fcc cell (172 atoms) using the DFT, while nanocubes up to $50\times50\times50$ (515\,151 atoms) were easily calculated using MD. 
A nanocube formed from $n\times n\times n$ conventional cubic fcc cells (4 atoms per cell) contains $N=4 n^3+6n^2+3n+1$ of atoms.
The calculated surface energy values shown in Fig.~\ref{fig:nanocubes} were fitted with an exponential relationship
\begin{equation}
  \gamma=\gamma_0 \exp\left(\frac{\cal A}{a}\right)\ ,\label{eq:fit}
\end{equation}
where $a=n\cdot a_0$ is the side length of a cube formed by $n\times n\times n$ conventional fcc cells with the lattice parameter $a_0$.
The quantities $\gamma_0$ and $\cal A$ are used as two fitting parameters.
The thus obtained values of the pre-exponential parameter, $\gamma_0^{\text{GGA}}=57.4\uu{meV/\AA^2}$, $\gamma_0^{\text{LDA}}=89.8\uu{meV/\AA^2}$, and  $\gamma_0^{\text{MD}}=81.4\uu{meV/\AA^2}$ agree well with the bulk surface energies for the $(1\,0\,0)$ facets ($\gamma_{(1\,0\,0)}^{\text{GGA}}=54.5\uu{meV/\AA^2}$, $\gamma_{(1\,0\,0)}^{\text{LDA}}=83.5\uu{meV/\AA^2}$, and $\gamma_{(1\,0\,0)}^{\text{MD}}=80.9\uu{meV/\AA^2}$). 
This is an expected result as the bulk values are limits for infinitely large cubes. 
It is, however, surprising, that such a good agreement is obtained for the DFT data where only three data points are available for the fitting procedure. 
The same fitting procedure yielded for the parameter ${\cal A}$ (Eq.~\ref{eq:fit}) values of $0.397\uu{nm}$, $0.392\uu{nm}$, and $0.661\uu{nm}$ for DFT-GGA, DFT-LDA, and MD data sets, respectively.

\subsubsection{Nanospheres}
\begin{figure}
  \centering
  \includegraphics[height=\figwidth]{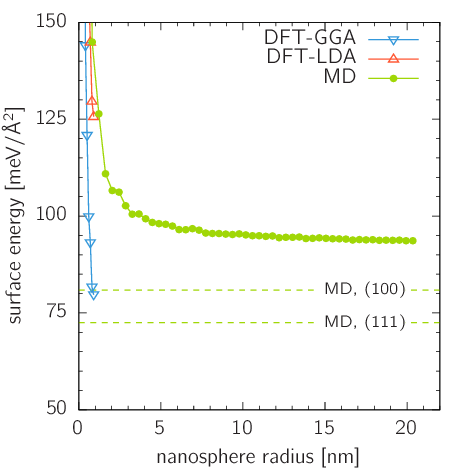}
  \caption{Surface energy of nanospheres calculated by DFT and MD.
    The dashed lines are the MD values for single-orientated $(1\,0\,0)$ and $(1\,1\,1)$ surfaces as listed in Table~\ref{tab:surface_energy}.}
  \label{fig:nanospheres}
\end{figure}

Nanospheres with all possible facet orientations were considered as an opposite extreme to the nanocubes with only a single orientation of their facets. 
They were constructed by cutting material contained in an ideal sphere of a given radius out of an infinitely large fcc Au crystal. 
The DFT calculations were performed up to $r=0.9\uu{nm}$ (152 atoms), while the MD calculations allowed easily for spheres up to $r=20.3\uu{nm}$ (2\,094\,177 atoms) (Fig.~\ref{fig:nanospheres}). 
In comparison to the case of nanocubes, the surface energy of the nanospheres converges faster to a constant value of $\approx94\uu{meV/\AA^2}$. 
This is a slightly higher value than $\gamma$ of any low-index facet (cnf.~Table~\ref{tab:surface_energy}) reflecting the fact that a spherical surface composes (from the atomistic point of view) of a large number differently orientated facets. 
Places where these facets meet (i.e., edges) are composed of atoms with the same or higher number of broken bonds than atoms in the surrounding planar facets, thus, further increasing the surface energy.

\subsubsection{Cuboctahedrons}

\begin{figure}
  \centering
  \setlength{\unitlength}{\figwidth}
  \begin{picture}(0,0)
    \put(0.55,0.55){\includegraphics[width=0.3\unitlength]{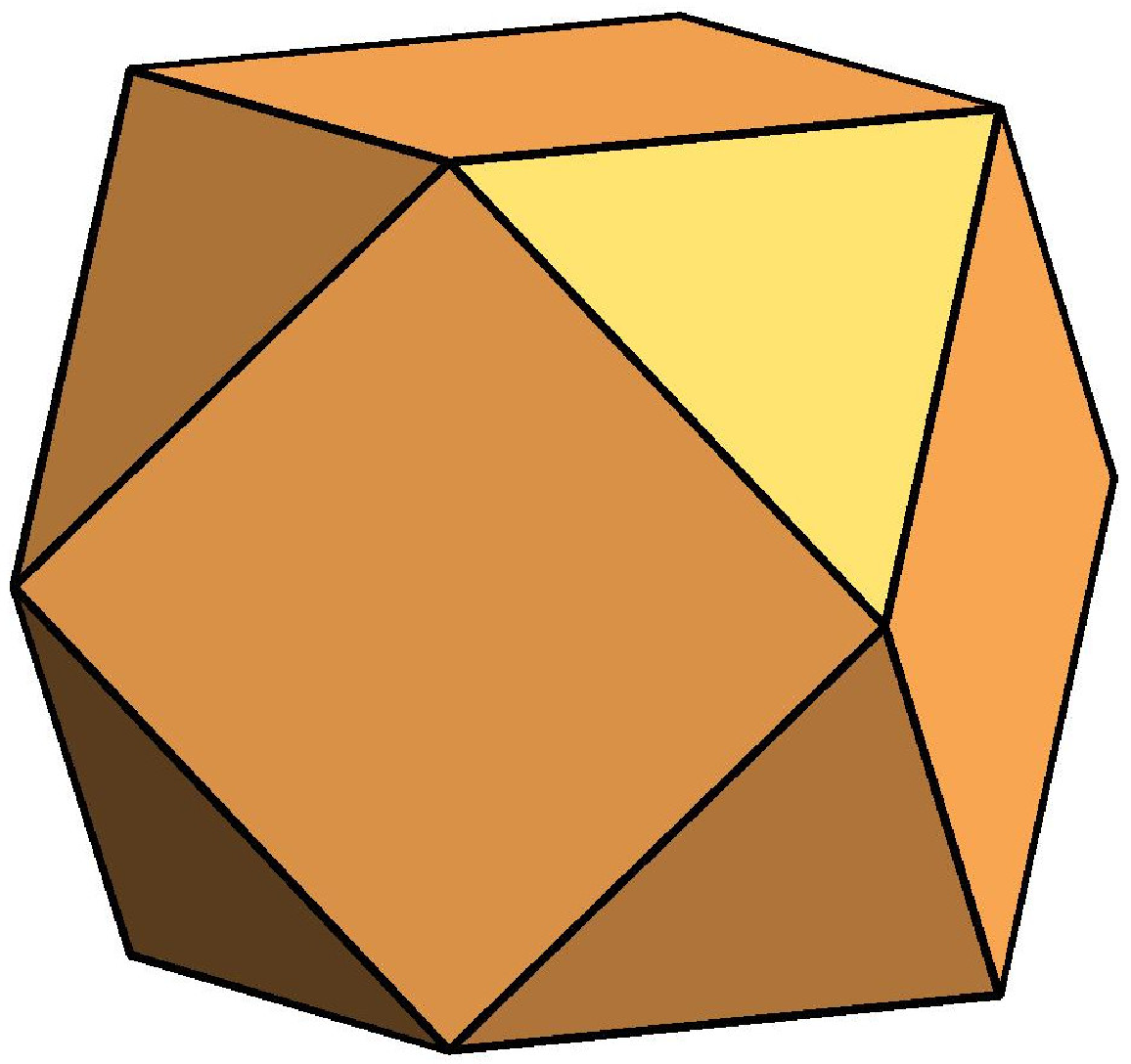}}
  \end{picture}
  \includegraphics[height=\figwidth]{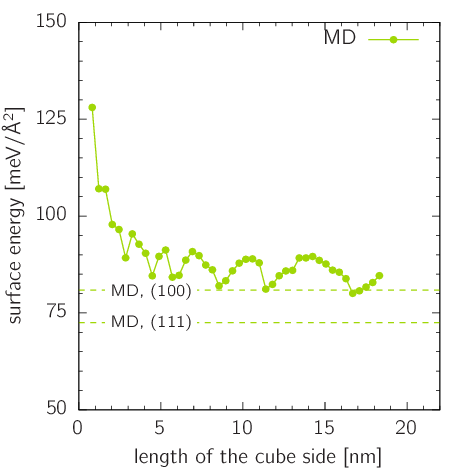}
  \caption{Surface energy of cuboctahedrons calculated by MD and showed as a function of the size of ``parent'' cube.
    The dashed lines are the MD values for single-orientated $(1\,0\,0)$ and $(1\,1\,1)$ surfaces as listed in Table~\ref{tab:surface_energy}.}
  \label{fig:cuboctahedrons}
\end{figure}

The last class of objects studied in this work are cuboctahedrons, i.e., $(1\,0\,0)$-faceted cubes with all apexes cut by $(1\,1\,1)$ planes (see inset in Fig.~\ref{fig:cuboctahedrons}). 
It can be seen that the surface energy oscillates between two values, $\approx 78$ and $\approx90\uu{meV/\AA^2}$. 
This behaviour is caused by the changing ratio of surface atoms forming the $(1\,0\,0)$ and $(1\,1\,1)$ facets and the edges and corners, which directly corresponds with the atomistic nature of the nanoparticle. 
A detailed analysis of the coordination of the surface atoms reveals that the number of 9-coordinated surface atoms, corresponding to ideal $(1\,1\,1)$ facets, is in anti-phase with the surface energy as shown in Fig.~\ref{fig:cuboctahedrons}. 
The 8-coordinated $(1\,0\,0)$ surface atoms also show small steps hence causing a non-monotonous increase of their number as a function of the cuboctahedron size. 
At the same time, the numbers of 10-, 7-, 6-, and 5-coordinated surface atoms forming edges and corners (i.e., atoms with even smaller coordination and, consequently, more broken bonds than those on ideal $(1\,0\,0)$ and $(1\,1\,1)$ facets, and hence increasing the overall surface energy), exhibit the same ``oscillations'' concerning the cuboctahedron size as the surface energy itself. 
Therefore, the oscillations are expected to decrease with increasing cuboctahedron size. 
It is interesting to note that the two limit values for the surface energies represent the same range as the two values, $80.9$ and $72.5\uu{meV/\AA^2}$ for pure $(1\,0\,0)$ and $(1\,1\,1)$ facets, respectively. 
Similarly to the case of nanospheres, the values are somewhat higher than the ideal single-orientated facets due to the presence of the edges and corners.

\section{Discussion}

\subsection{Correction of the surface area for electronic cloud}

The surface areas calculated in the previous parts corresponds to the convex hull of ionic positions.
In our recent paper~\cite{Holec2017-qy} dealing with predicting surface energy of Au$_{55}$ cluster, we have discussed the error made by neglecting extend of the electronic cloud.
There, a radius correction of $1.3$--$1.4\uu{\AA}$ has been proposed under the assumption that the mass density of the nanocluster is the same as that of bulk fcc-Au.
On the other hand, radius corrections of $0.5$--$0.8\uu{\AA}$ have been proposed by \citet{De_Heer1993-ue}.

In order to see how neglecting the electronic cloud layer actually influences the predicted surface energies, we re-evaluate the surface areas. 
Let $\{\vec R_i\}$ be a set of the atomic (ionic) positions defined with respect to the nanoparticle centre of mass, i.e.
\begin{equation}
  \sum_i \vec R_i=\vec 0\ ,
\end{equation}
where the sum is performed over all atoms in the nanoparticle.
Subsequently, a new set of coordinates, $\{\vec{\tilde R}_i\}$, is defined as
\begin{equation}
  \vec{\tilde R}_i=\left(|\vec R_i|+\Delta\right)\vec R_i^0
\end{equation}
where $\vec R_i^0=\vec R_i/|\vec R_i|$ is a unit vector along the direction of $\vec R_i$.
This means that all atoms, and in particular those on the convex hull envelope, are shifted by $\Delta$ away from the nanoparticle centre of mass.
A new surface area is calculated as a convex hull of $\{\vec{\tilde R}_i\}$ positions for several representative values of $\Delta$.

\begin{figure}[ht]
  \centering
  \begin{subfigure}[t]{\smallfigwidth}
    \caption{Nanocubes}
    \includegraphics[height=\smallfigwidth]{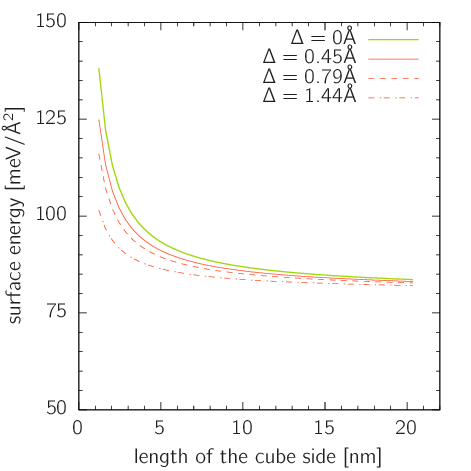}
    \label{fig:nanocubes_corrected}
  \end{subfigure}
  \begin{subfigure}[t]{\smallfigwidth}
    \caption{Cuboctahedrons}
    \includegraphics[height=\smallfigwidth]{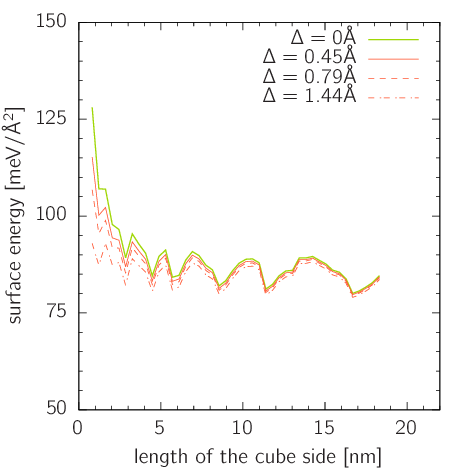}
    \label{fig:cuboctahedrons_corrected}
  \end{subfigure}
  \begin{subfigure}[t]{\smallfigwidth}
    \caption{Nanospheres}
    \includegraphics[height=\smallfigwidth]{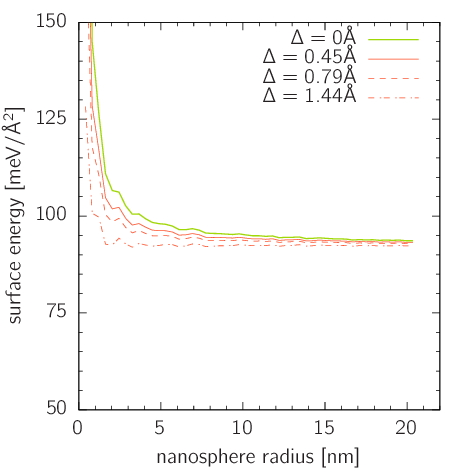}
    \label{fig:nanospheres_corrected}
  \end{subfigure}
  
  \begin{subfigure}[t]{\smallfigwidth}
    \caption{Nanocubes}
    \includegraphics[height=\smallfigwidth]{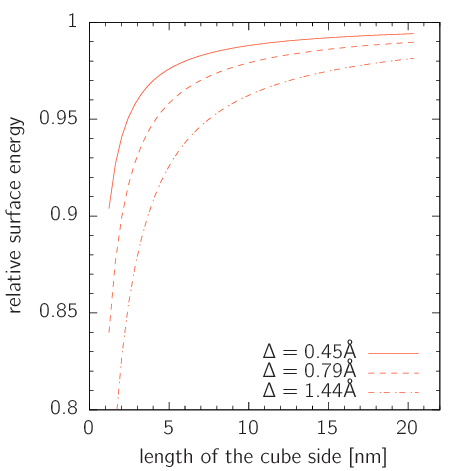}
    \label{fig:nanocubes_corrected_rel}
  \end{subfigure}
  \begin{subfigure}[t]{\smallfigwidth}
    \caption{Cuboctahedrons}
    \includegraphics[height=\smallfigwidth]{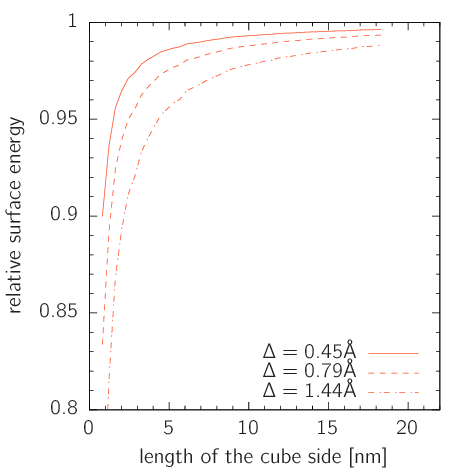}
    \label{fig:cuboctahedrons_corrected_rel}
  \end{subfigure}
  \begin{subfigure}[t]{\smallfigwidth}
    \caption{Nanospheres}
    \includegraphics[height=\smallfigwidth]{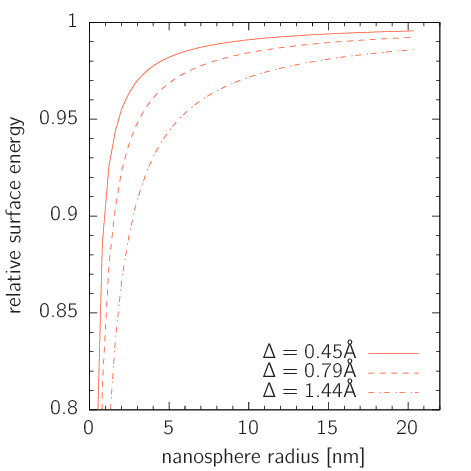}
    \label{fig:nanospheres_corrected_rel}
  \end{subfigure}
  \caption{Corrected absolute (upper row) and relative values (lower row) of the surface energies for (a), (d) nanocubes, (b), (e) cuboctahedrons, and (c), (f) nanosheres.
    The relative surface energies are calculated with respect to the values without correction for the electronic cloud thickness ($\Delta=0$).
    \label{fig:gamma_corrected}}
\end{figure}

The results are summarised in Fig.~\ref{fig:gamma_corrected} for all three nanoparticle geometries considered in the present work.
In all cases, the surface energy decreases with increasing values of $\Delta$, which is simply a consequence of the surface energy definition in Eq.~\ref{eq:gamma}.
It is, however, remarkable to notice that even for the largest nanoparticle sizes the surface energy reduction is still larger than $1\%$ for the DFT-based electron cloud thickness.
We, therefore, conclude that, especially for nanoparticles with specific sizes below $5\uu{nm}$, the correction of the surface area due to the electronic cloud is essential.
Moreover, it is likely that for the small nanoparticle sizes, the surface energies calculated here overestimated due to that fact that even lower energy can be obtained for other atomic ordering than fcc (e.g., Mackay icosahedrons as in the case of Au$_{55}$) or even amorphous liquid-like structures \cite{Holec2017-qy}.
Finally, it is worth noting that the problem of electronic cloud is not an issue in standard calculations of single orientated flat single crystal facets since it does not influence the actual surface area.

\subsection{Surface induced excess energy}\label{sec:excess_energy}
\begin{figure}
  \centering
  \begin{subfigure}[t]{\figwidth}
    \caption{}
    \includegraphics[height=\figwidth]{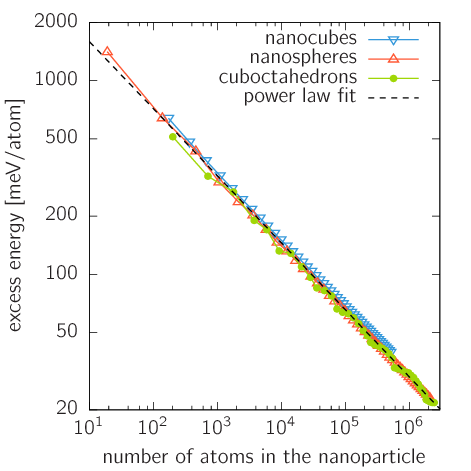}
    \label{fig:excess_energy}
  \end{subfigure}
  \begin{subfigure}[t]{\figwidth}
    \caption{}
    \includegraphics[height=\figwidth]{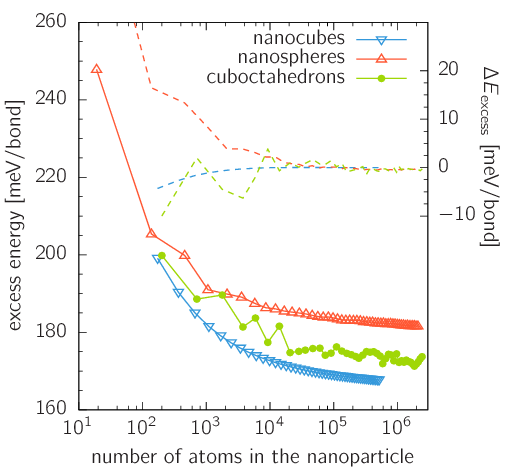}
    \label{fig:excess_energy_bond}
  \end{subfigure}
  \caption{Excess energy, $E_{\text{excess}}$, of nanoparticles with respect to the bulk fcc Au as a function of the number, $N$, of atoms forming the nanoobject. 
           $E_{\text{excess}}$ is normalised to (a) number of the atoms forming the nanoparticle, and (b) to the number of broken bonds.
           The dashed lines in (b) show the difference between the actual value of $E_{\text{excess}}$ as calculated by MD, and a fitted value using Eq.~\ref{eq:excess_general}.}
\end{figure}

As mentioned above and discussed in the literature, the surace area of nanoparticles is an ill-defined quantity.
In order to eliminate this problem, we introduce a new quantity $E_{\text{excess}}$ expressing the surface-induced excess energy with respect to the bulk energy corresponding to the same number, $N$, of atoms as in the nanoparticle, normalised to 1 atom, as
\begin{equation}
  E_{\text{excess}}=\frac{E_{\text{nanoparticle}}-NE_{\text{fcc-Au}}}{N}\ .
\end{equation}
A similar concept has been previously demonstrated to work also for energetics of carbon fullerenes \cite{Holec2010-dr}, or even for elasticity of nanoporous gold \cite{Mameka2014-as}. 
If the excess energy, $E_{\text{excess}}$, is evaluated for nanocubes, nanospheres, and cuboctahedrons, a linear relationship between $\log E_{\text{excess}}$ and $\log N$ is obtained independent of the nanoparticle shape (Fig.~\ref{fig:excess_energy}). 
This suggests that the excess energy is a power law function of the total number of atoms (nanoparticle size). 
This fit (the dashed line in Fig.~\ref{fig:excess_energy}) gives
\begin{equation}
  E_{\text{excess}}=3523.3\uu{meV/atom}\times N^{-0.346}\ .
\end{equation}

Recalling the idea that the surface energy is genuinely connected with the broken bonds (bb), we now establish the energy needed to ``break'' a bond. 
Let us consider an $n\times n\times n$ nanocube containing atoms with 4 different nearest neighbour coordinations: 8 atoms with 9 bb forming corners (i.e.,  3-coordinated atoms), $(12n-12)$ atoms with 7 bb forming the edges (i.e., 5-coordinated atoms), $(12n^2-12n+6)$ atoms with 4 bb forming the surface facets (i.e., 8-coordinated atoms), and $(4n^3-6n^2+3n-1)$ bulk atoms with no bb (i.e., fully 12-coordinated atoms). 
If we simply assume that all bonds ``cost'' the the same energy $E_{\text{bond}}$ to break them, the excess energy follows as a sum of the contributions described above, as
\begin{equation}
   E_{\text{excess}}=\left[9\times8+7\times(12n-12)+4\times(12n^2-12n+6)\right]E_{\text{bond}}\ ,
   \label{eq:simple_model}
\end{equation}
yielding $E_{\text{bond}}=168.1\uu{meV/bond}$ from fitting the nanocubes data.

However, the red triangles in Fig.~\ref{fig:excess_energy_bond}, showing the nanocubes excess energy normalised to the number of broken bonds, clearly exhibit a non-constant value for $E_{\text{bond}}$. 
Consequently, we propose a slightly modified description in which the energy needed to break a bond is a function of the coordination. 
Hence, it costs different energy to create, e.g., a corner atom (9 broken bonds) than a facet atom (4 broken bonds). 
Thus the excess energy becomes
\begin{equation}
  E_{\text{excess}}=72 E_{\text{corner}}+84(n-1)E_{\text{edge}}+24(2n^2-2n+1)E_{\text{facet}}\ .
  \label{eq:excess_energy}
\end{equation}
Fitted yields $E_{\text{corner}}=272.1\uu{meV/bond}$, $E_{\text{edge}}=215.2\uu{meV/bond}$, and $E_{\text{facet}}=166.0\uu{meV/bond}$. 
The fitted values of the excess energy, normalised to the number of bonds, are plotted in Fig.~\ref{fig:excess_energy_bond} with the red solid line. 
It turns out that for nanocubes with side $\gtrapprox5\uu{nm}$, Eq.~\ref{eq:excess_energy} provides predictions with an accuracy better than $\approx1\uu{meV/bond}$. Energy of a broken bond corresponding to an infinitely large $(1\,0\,0)$ facet can be estimated from the surface energies as given in Table~\ref{tab:surface_energy}. 
This value is $167.5\uu{meV/bond}$, which is close to $E_{\text{bond}}$, Eq.~\ref{eq:simple_model}, as well as $E_{\text{facet}}$ (Eq.~\ref{eq:excess_energy}).

The complex shapes of cuboctahedrons and nanospheres somewhat restrict the intuitive analysis of the excess energy above presented. 
When the excess energy is fitted with a single valued energy per broken bond (equivalent to Eq.~\ref{eq:simple_model}), values of $172.8\uu{meV/bond}$ and $181.9\uu{meV/bond}$ are obtained for cuboctahedrons and nanospheres, respectively. 
These values represent an excellent estimation of the excess energies in the limit of large nanoparticles, as shown in Fig.~\ref{fig:excess_energy_bond}. 
Moreover, the excess energy value for cuboctahedrons lies between the values estimated for $(1\,0\,0)$ ($E_{(1\,0\,0)}=167.5\uu{meV/bond}$) and $(1\,1\,1)$ ($E_{(1\,1\,1)}=173.3\uu{meV/bond}$) facets. 
This further illustrates that the surface energy values, as presented in Sec.~\ref{sec:surface_energy}, are remarkably influenced by the evaluation of the actual surface area (which is, from the atomistic point of view, ill-defined). 
Consequently, the mean value  of the surface energy of cuboctahedrons as shown in Fig.~\ref{fig:cuboctahedrons} lies outside the range bounded by $\gamma_{(1\,0\,0)}$ and  $\gamma_{(1\,1\,1)}$ values.

Finally, in order to obtain a non-constant behaviour, we fit the excess energy with
\begin{equation}
  E_{\text{excess}}=\sum_{i=1}^{11} (12-i) N(i) E(i)\label{eq:excess_general}
\end{equation}
where $N(i)$ is the number of $i$-coordinated atoms (i.e. those having $(12-i)$ broken bonds) and $E(i)$ is the corresponding excess energy contribution. 
Eq.~\ref{eq:excess_general} is a generalised formulation of Eq.~\ref{eq:excess_energy} reflecting that all possible coordinations may occur due to the shape of nanoparticles. 
We note that the smallest coordination obtained was 3 and 4 for the case of cuboctahedrons and nanospheres, respectively. 
The fitted values of $E(i)$ are given in Table~\ref{tab:Ei_fitted}, and the difference between the actual $E_{\text{excess}}$ from MD and values predicted using Eq.~\ref{eq:excess_general} is shown in Fig.~\ref{fig:excess_energy_bond} with dashed lines. 
Obviously, the fit provides excellent agreement for nanoparticles containing $\approx10^4$ atoms and more.

\begin{table*}
\centering
\begin{tabular}{l|rrr}
  & nanocubes & cuboctahedrons & nanospheres \\\hline
  $E(3)$ [meV/bond]& 272.1 & 287.3 & 0 \\
  $E(4)$ [meV/bond]& 0 & 161.1 & 426.3 \\
  $E(5)$ [meV/bond]& 215.2 & 243.4 & 258.3 \\
  $E(6)$ [meV/bond]& 0 & 163.1 & 232.0 \\
  $E(7)$ [meV/bond]& 0 & 239.5 & 212.2 \\
  $E(8)$ [meV/bond]& 166.0 & 170.3 & 181.1 \\
  $E(9)$ [meV/bond]& 0 & 162.2 & 159.2 \\
  $E(10)$ [meV/bond]& 0 & 93.6 & 100.7 \\
  $E(11)$ [meV/bond]& 0 & 16.9 & 46.0
\end{tabular}
\caption{Fitted coefficients $E(i)$ for the excess energy expression according to Eq.~\ref{eq:excess_general}.
         The index $i$ expresses the coordination of atoms (i.e., $12-i$ is the number of bb).}
\label{tab:Ei_fitted}
\end{table*}

Our analysis provides an insight into the here predicted trends. 
Regardless of the nanoparticle shape, the surface energy decreases with the increasing particle size. 
The reason is that the smaller is the nanoparticle, the larger is the fraction of the surface atoms with small coordination, i.e., those with lots of broken bonds. Moreover, the energy to break a bond increases (generally) with the decreasing atom coordination.

\subsection{Contribution of surface stress state}

As it has been recently stressed out \cite{Muller2014-ih}, the excess energy due to a free surface has two contributions: the surface energy contribution related to the energy penalty of broken bond and the contribution due to the elastic strain energy generated by the surface stress state. 
The latter depends on the surface curvature. 
As an illustrative example let us assume a spherical body and a homogeneous surface stress state with the value $\sigma$ acting on it, which leads to a pressure with value $2\sigma/R$ in the whole spherical body. 
From this description it becomes clear that the energetic surface stress contribution is zero for the slab approach. 
Similarly, the energetic surface stress contribution will be negligible for rather large nanocubes with only a marginal fraction of corner and edge atoms (see discussion in the section~\ref{sec:excess_energy}).

We now try to estimate the energetic surface stress contribution to the excess energy for the case of a spherical nanoparticle using classical continuum mechanics. 
Let us denote $R$ the nanosphere's radius, and $\gamma$ its surface energy. 
Furthermore, let us keep to the reasonable assumption that the value of $\sigma$ and $\gamma$ are of the same order of magnitude. 
The corresponding total surface energy is then
\begin{equation}
  {\cal E}_{\gamma}=4\pi R^2\gamma\ .
\end{equation}
For sake of simplicity we further assume isotropic elastic properties of the nanoparticle, with $\nu$ and $E$ being its Poisson's ratio and Young's modulus, respectively. 
The elastic strain energy caused by the surface stress $\sigma$, activating an internal pressure $2\sigma/R$, is
\begin{equation}
 {\cal E}_{\sigma}=\frac43\pi R^3\frac{6(1-2\nu)}{E}\frac{\sigma^2}{R^2}\ ,
\end{equation}
for details see, e.g, Ref.~\onlinecite{Fischer2008-tj}, Appendix 3. 
The ratio of the energetic surface stress contribution to the surface energy follows with $\sigma=\gamma$ as
\begin{equation}
 \frac{{\cal E}_{\sigma}}{{\cal E}_{\gamma}}=\frac{2(1-2\nu)}{E}\frac{\gamma}{R}\ .\label{eq:ratio}
\end{equation}
Taking a representative values for gold, $\gamma=1\uu{J/m}^2$, $E=78\uu{GPa}$, $\nu=0.44$, and $R=1\uu{nm}$, Eq.~\ref{eq:ratio} yields $0.359\times10^{-2}$, i.e. the energetic surface stress contribution to the total excess energy is less than $1\,\%$ of the surface induced excess energy. 
This ratio becomes even smaller (negligible) for larger nanospheres.

\begin{figure*}
  \centering
  \begin{subfigure}[t]{\figwidth}
    \caption{}
    \includegraphics[height=\figwidth]{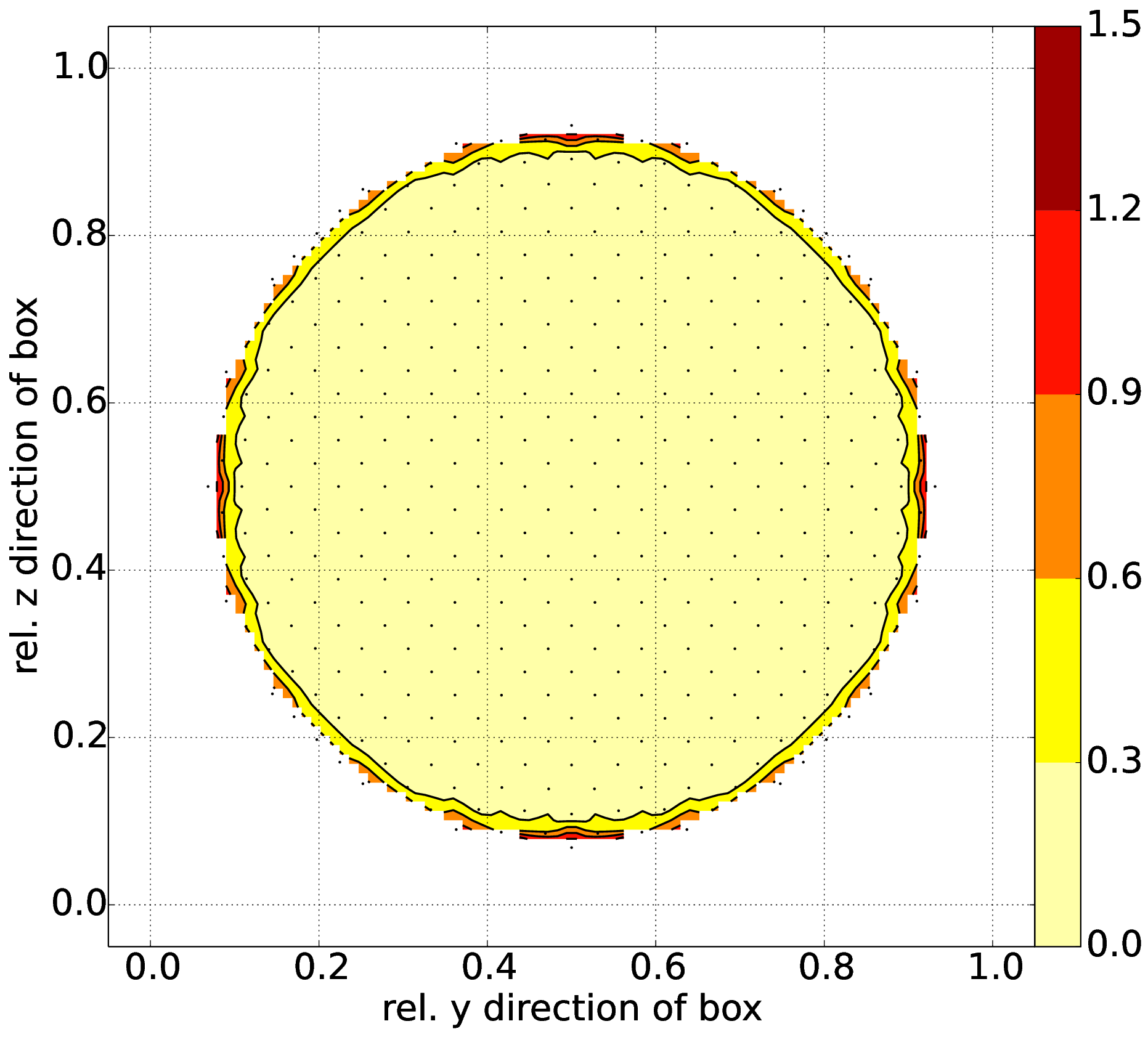}
    \label{fig:excess_energy_nanosphere}
  \end{subfigure}
  \begin{subfigure}[t]{\figwidth}
    \caption{}
    \includegraphics[height=\figwidth]{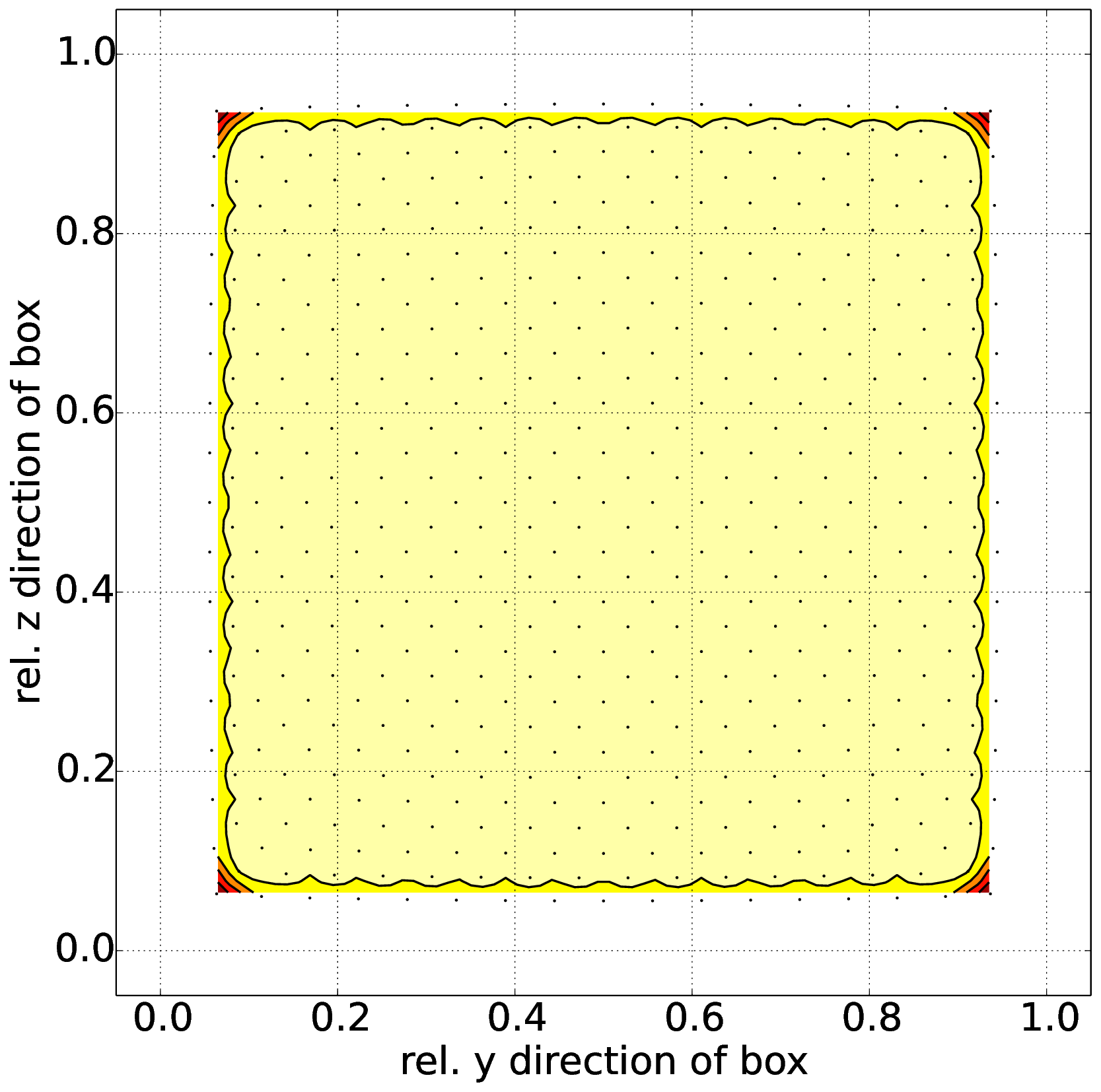}
    \label{fig:excess_energy_nanocube}
  \end{subfigure}
  \caption{Contour plots of the distribution of the surface stress induced excess energy (per atom) contribution for a cross section of (a) a nanoshere ($R=3.25\uu{nm}$) and (b) a nanocube ($a=6.92\uu{nm}$). 
    Both cross sections include the nanoparticle centre. 
    The dots represent actual atoms in the cross section, e.g. real locations, where the excess energy is stored. 
    For sake of clear demonstration, the discrete data were interpolated over the whole cross sectional area.
    \label{fig:excess_energy_contour}}
\end{figure*}

To corroborate this rather simplistic estimation, we plot the excess energy distribution over a cross section including the centre for a nanosphere (Fig.~\ref{fig:excess_energy_nanosphere}) and a nanocube (Fig.~\ref{fig:excess_energy_nanocube}) as obtained from the MD simulations. 
Several observations can be made. 
Firstly, the excess energy is concentrated at the nanoparticle surface irrespective of its shape. 
The surface stress (and hence the corresponding elastic strain energy) should be only of relevance for a nanosphere.
However, we can conclude that this contribution is effectively zero (or negligible).
A similar situation can be expected for a nanocube, where the excess energy is concentrated to the nanocube edges (corner of the cross section in Fig.~\ref{fig:excess_energy_nanocube}). 
This fact nicely agrees with the fitted values of $E_{\text{edge}}=215.2\uu{meV/bond}$ being larger than $E_{\text{facet}}=166.0\uu{meV/bond}$, estimated in section \ref{sec:excess_energy}.

Even though the term surface energy was used in a slightly imprecise way throughout the section \ref{sec:surface_energy} (more accurate would be to talk about surface induced excess energy), we conclude that the energy contribution of surface stress can be neglected and the two quantities, surface energy and surface induced excess energy, are equivalent for practical cases with nanoparticles larger than $\approx 1\uu{nm}$.

\section{Conclusions}

A molecular dynamics study, complemented by first principles Density Functional Theory calculations, was performed to obtain surface energy of small gold nanoclusters of various sizes and (geometrically well defined) shapes. 
The employed interatomic pair potential was shown to give structural parameters and surface energies comparable with DFT-LDA calculations. 
The surface energy of nanocubes and nanospheres has been shown to converge to a constant value. 
The convergence was faster in the case of nanospheres compared with nanocubes.
The surface energy, $\gamma$, is practically constant for any particles with radius larger than $\approx 3\uu{nm}$. 
Truncated cubes (cuboctahedrons) did not achieve a single value for the surface energy within the studied range of nanoparticle sizes but, instead, an oscillating behaviour between two values. 
The range of these oscillations equals to the difference between $\gamma$ of $(1\,0\,0)$ and $(1\,1\,1)$ facets. 
Finally, the surface-induced excess energy obviously follows a universal power-law dependence on the number of atoms forming the nanoparticle and is, to a large extent, related to a number of broken bonds (reduced coordination of the surface atoms). 
Importantly, the size-dependence of surface energy becomes significantly reduced when the actual surface area is corrected by the thickness of the electronic cloud, leading to almost constant values particularly for nanocube and nanosphere sizes of about $5\uu{nm}$ and more.

As outlined above, this study has found an increase of the surface energy with decreasing particle size (which is in agreement with other theoretical studies). 
Two remarks may be useful in this regard.
Firstly, this fact should not be confused with experimental works on liquid solution--solid nanoparticle interface energies of gold nanoparticles, moreover often having irregular shapes or even liquid-like surface layer. 
Secondly, we note that small nanoparticles, specifically the Au$_{55}$, were shown to be amorphous rather than crystalline.
Hence the values predicted here for the smallest particle sizes of a few nanometers are not     relevant for amorphous or glassy particles.

In conclusion, this work contributes to understanding of surface energy (solid phase--vacuum interface) of crystalline mesoparticles and its relation to the their structure.

\section*{Acknowledgements}
The computational results presented have been achieved [in part] using the Vienna Scientific Cluster (VSC).

\section*{References}

\end{document}